 \documentclass[final,1p,times]{elsarticle}
\usepackage{graphicx}
\usepackage{amssymb}
\begin{document}

\begin{frontmatter}

\title{2MASS photometry and kinematical studies of open cluster NGC 188}

\author{W. H. Elsanhoury$^a,^b$\footnote{E-mail:  welsanhoury@gmail.com (W. H. Elsanhoury); aaharoon@kau.edu.sa (A. A. Haroon); chupina@inasan.ru~(N.~V.~Chupina); svvs@ya.ru~(S.~V.~Vereshchagin); deveshpath@gmail.com~(Devesh P. Sariya); rkant@aries.res.in (R. K. S. Yadav); jiang@phys.nthu.edu.tw (Ing-Guey Jiang)}, A. A. Haroon$^a,^c$, N. V. Chupina$^d$, S. V. Vereshchagin$^d$, Devesh P. Sariya$^e$\footnote{Corresponding author; Tel:+886-3-571-5131 ext. 42666; Fax No: +886-3-572-3053}, R. K. S. Yadav$^f$, Ing-Guey Jiang$^e$}

\address{$^{a}$Astronomy Department, National Research Institute of Astronomy and Geophysics (NRIAG) 11421, Helwan, Cairo, Egypt\\
$^{b}$Physics Department, Faculty of Science, Northern Border University, Rafha Branch, Saudi Arabia\\
$^{c}$Astronomy Department, Faculty of Science, King Abdul Aziz University, Jeddah, Saudi Arabia\\
$^{d}$Institute of Astronomy Russian Academy of Sciences (INASAN),48 Pyatnitskaya st., Moscow, Russia\\
$^{e}$Department of Physics and Institute of Astronomy, National Tsing Hua University, Hsin-Chu, Taiwan\\
$^{f}$Aryabhatta Research Institute of observational sciencES (ARIES), Manora Peak Nainital 263 002, India.}
\begin{abstract}
%% Text of abstract
In this paper, we present our results for the photometric and kinematical studies of old open cluster NGC 188.  
We determined various astrophysical parameters like limited radius, core and tidal radii, distance,
luminosity and mass functions, total mass, relaxation time etc. for the cluster using 2MASS catalog.
We obtained the cluster's distance from the Sun as 1721$\pm$41 pc and log (age)= 9.85$\pm$0.05 at Solar metallicity. 
The relaxation time of the cluster is smaller than the estimated cluster age which suggests that the 
cluster is dynamically relaxed. Our results agree with the values mentioned in the literature.
We also determined the cluster’s apex coordinates as ($281^\circ$.88, $-44^\circ$.76) using AD-diagram method.  
Other kinematical parameters like space velocity components, cluster center 
and elements of Solar motion etc. have also been computed.

\end{abstract}


\end{frontmatter}

\section{Introduction}
\label{Intro}
NGC 188 is one of the oldest known open clusters of the Milky Way
belonging to the old Galactic disk population (Platais et al. 2003).
The cluster's spatial location in the Galaxy
($l=122^\circ$.8, $b=22^\circ$.4) makes it relatively
less obscured and poorly contaminated by field stars, hence, being easy
to observe (Meibom et al. 2009; Wang et al. 2015; von Hippel \& Sarajedini 1998). 
Thus, NGC 188 has been used extensively as a classical reference to study the
evolution of Galactic disk and chemical and dynamical evolution
of the Galaxy. 
A plot of the Galactic orbit of NGC 188 is given by
Carraro \& Chiosi (1994).
One of the reasons behind the survival of NGC 188 to this advance age
may be its almost circular, highly inclined external orbit,
which avoids the inner disk regions for most of the times
(Bonatto et al. 2005).
The cluster is known to have approximately Solar metallicity
(Friel et al. 2002; Randich et al. 2003; Worthey \& Jowett 2003).
The color-magnitude diagram of the cluster shows a
wide giant branch (Twarog 1997; Norris \& Smith 1985),
blue stragglers (Leonard \& Linnel 1992; Dinescu et al. 1996)
and various type of binaries (von Hippel \& Sarajedini 1998).
Knowledge of accurate fundamental parameters of the open clusters
is essential for many astrophysical calibrations
(Fornal et al. 2007). The age of NGC 188 remains a crucial
factor and surveys like 2MASS (Skrutskie et al. 1997) have enabled
such studies in the near-IR bands. 

Open clusters are group of stars moving in parallel direction 
in space on the celestial sphere.
The directions of proper motions of the stars of open clusters will
converge to a point called vertex ($A_0,D_0$) of the cluster.
Proper motions can be used to calculate membership probabilities and
taking the most probable cluster members
for studying the kinematics of
the cluster will provide more precise determination of the vertex.

In light of the above discussion, we studied some of the 
basic parameters and kinematical properties of the cluster using the 2MASS catalogue in combination with 
other source of data which include information of proper motion and radial velocities of the cluster members. 
Conclusively, many parameters like limiting radius, core radius, age, distance,
mass function slope and dynamical relaxation time etc. could be estimated. 
Our kinematical studies of the cluster provided the convergent point of 
the cluster, cluster's velocity and the components of
space motion of the cluster. We also derived the elements of Solar motion
with respect to NGC 188. The kinematical parameters derived by us
will help in understanding the motion of the cluster in the Galaxy.
 
The structure of this article is as follows:
in Sect.~\ref{OBS}, we describe the data used for the present study.
Section~\ref{phot} explains the photometric study and the results achieved.
Section~\ref{kin} deals with the kinematical analysis of NGC 188.
The conclusions of this paper have been given in Sect.~\ref{con}.

\section{Data used and reduction procedures}
\label{OBS}

Due to the unavailability of all the required data in one general catalogue for the present study, we have
compiled data from different sources. We have taken photometric data from 2MASS catalogue (Cutri et al. 2003).
2MASS uniformly scanned the entire sky in three near-infrared bands $(J,H,K_s)$ to detect
and characterize point sources brighter than 1 mJy in each band, with signal-to-noise ratio (SNR) 
greater than 10, using a pixel size of 2$''$.0.

We combined the above photometric data with kinematical data for proper motions and radial velocities.
Platais et al. (2003) presented a technique to obtain precise proper motions of stars in the region of NGC 188,
using old photographic plates with assorted large aperture reflectors, in combination 
with recent CCD Mosaic Imager frames. They used their proper motions to determine astrometric 
membership probabilities down to $V$=21 in the 0.75 $deg^2$ area around NGC 188.
For our study, we consider stars with membership probabilities higher 
than 50 \% from the catalogue given by Platais et al. (2003), which gives us 562 member stars.
Geller et al. (2008) presented the results of ongoing radial velocity (RV) survey for NGC 188
by WIYN\footnote {Joint facility of the University of Wisconsin-Madison, 
Indiana University, Yale University, and the National Optical Astronomy Observatory}.
The data set observed by 3.5-m WIYN telescope on Kitt Peak in Arizona spans a 
time baseline of 11 years, a magnitude range of 12$\le V \le$16.5 (1.18-0.94 $M_{\odot}$)
and covers an area of one degree diameter on the sky. 
For the kinematical studies, the values of radial velocity for NGC 188 stars
have been taken from Geller et al. (2008).

\section {Photometric analysis}
\label{phot}
\subsection{Radial density profile (RDP)}

To determine radial extent of the cluster, we extracted $J,H,K_s$ magnitudes, positions  and radial 
distance of stars from the cluster centre. This data set of the cluster is taken from the Vizier 
webpage. The RDP 
were built by calculating the mean surface density in concentric rings
around the cluster center, 
$(\alpha, \delta)_{J2000}$=(00$^{\rm h}$47$^{\rm m}$13$^{\rm s}$.12, $85^\circ$14$'$51$''$).
We calculated the mean 
surface density $\rho(r)$ of each ring using King model (King 1966), i.e.:
\begin{center}
$\rho(r) = f_{bg} + \frac{f_0}{1+ (\frac{r}{r_{core}})^2} $
\end{center}
where $f_{bg}$ is the background surface density and $r_{core}$ is the core radius of the cluster where 
the stellar density, $\rho(r)$, becomes half of its central value, $\rho_0$.

In addition, we can also define the limited radius of the cluster ($r_{lim}$) which represents the radius 
which covers the entire cluster area and reaches enough stability with the background field density 
(Tadross $\&$ Bendary 2014). Mathematically,  $r_{lim}$ is defined as:
\begin{center}
$r_{lim} =  r_{core}\sqrt{\frac{f_0}{3\sigma_{bg}}-1} $
\end{center}

\begin{figure*}[]
\centering
\vspace{-0.7cm}
\includegraphics[width=8cm,height=8cm]{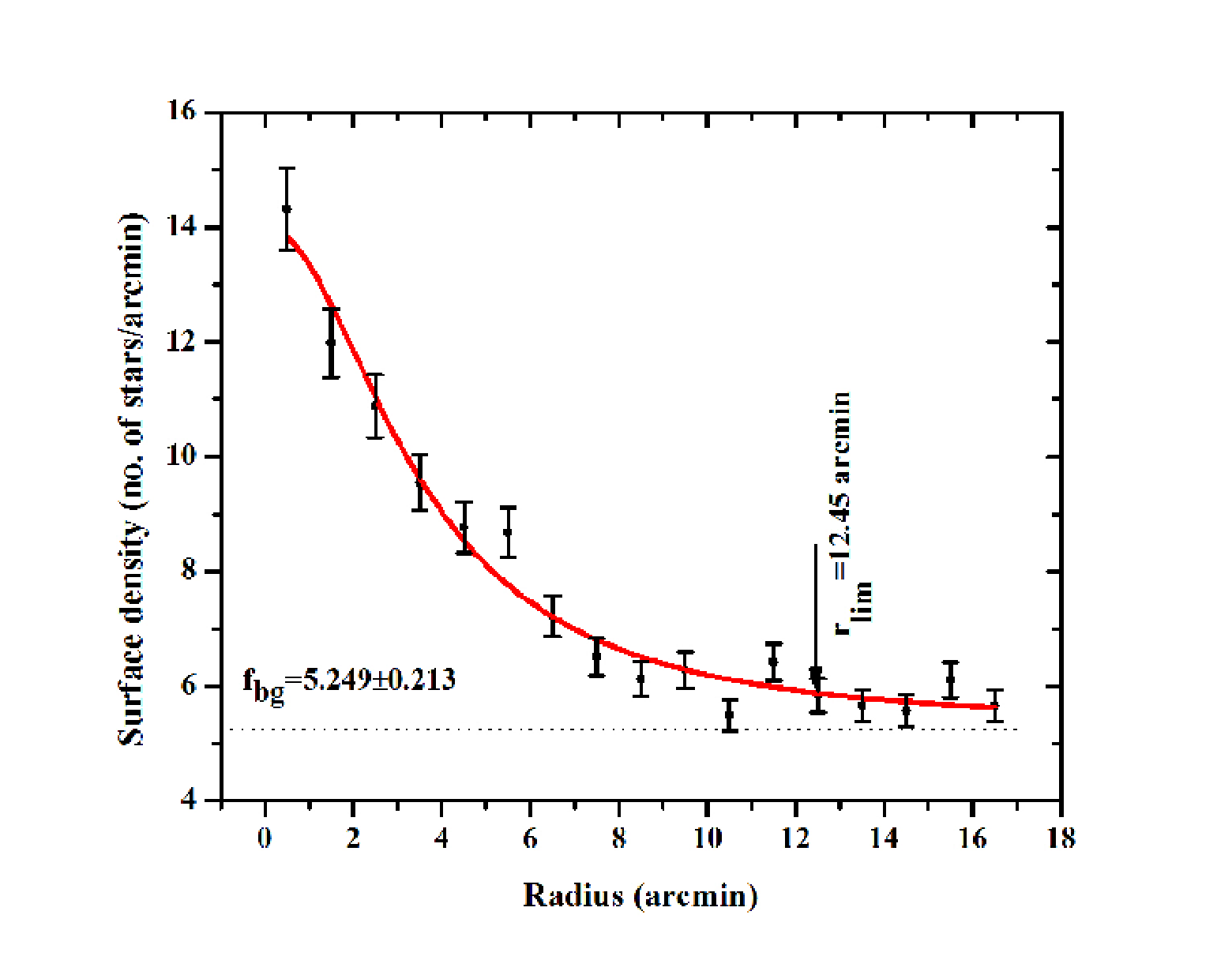}
\caption{The radial density profile of NGC 188. Solid line denotes the fitted density 
distribution and the dashed line represents the background density.}
\label{rdp}
\end{figure*}

Figure~\ref{rdp} represents the RDP of the cluster. The background stellar density is also shown
with the dotted line. Solid line represent the fitted profile of the cluster. Different radii of 
the clusters are derived and values are listed in Table 1. Our derived values are very similar 
to the values given in the literature.

\subsection{Color magnitude diagram (CMD) and isochrone fitting}

\begin{figure}[]
\centering
\vspace{-0.7cm}
\includegraphics [width=6cm]{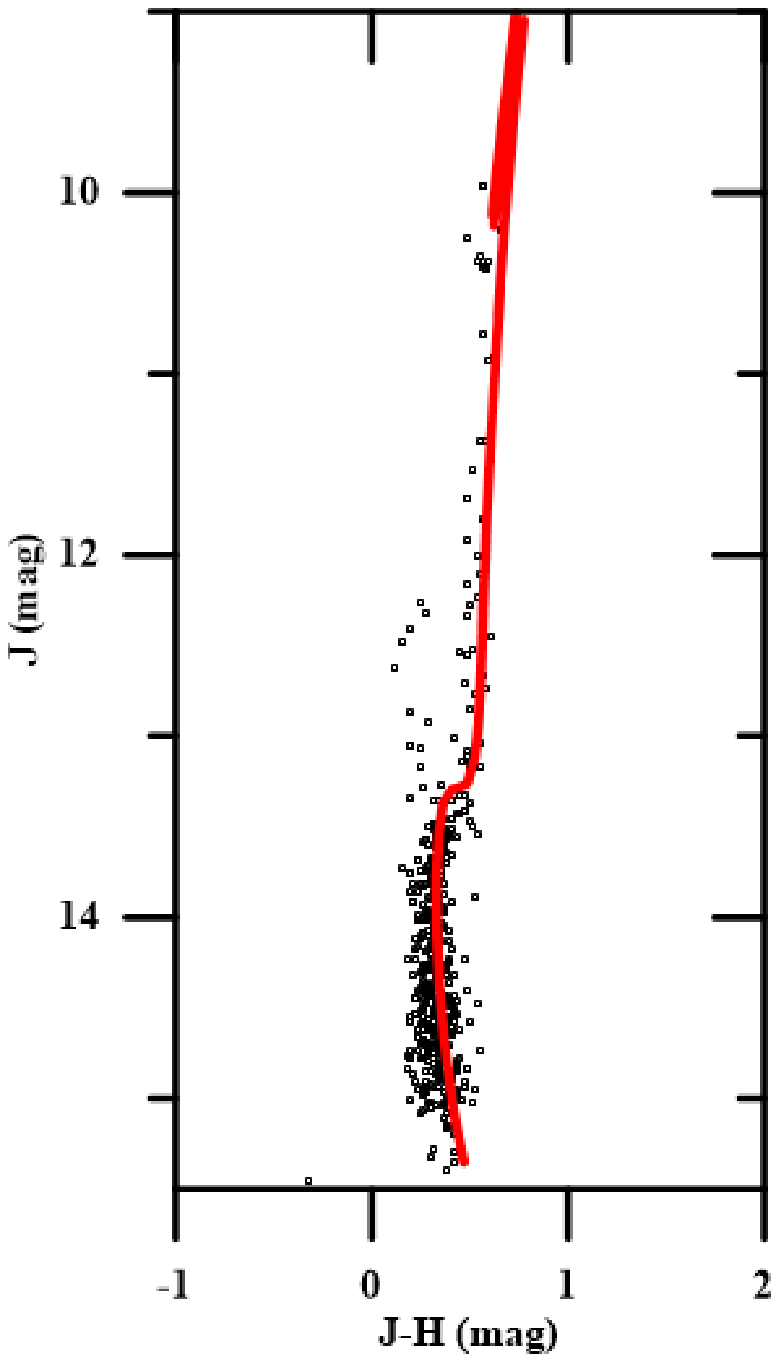} 
\includegraphics [width=6cm]{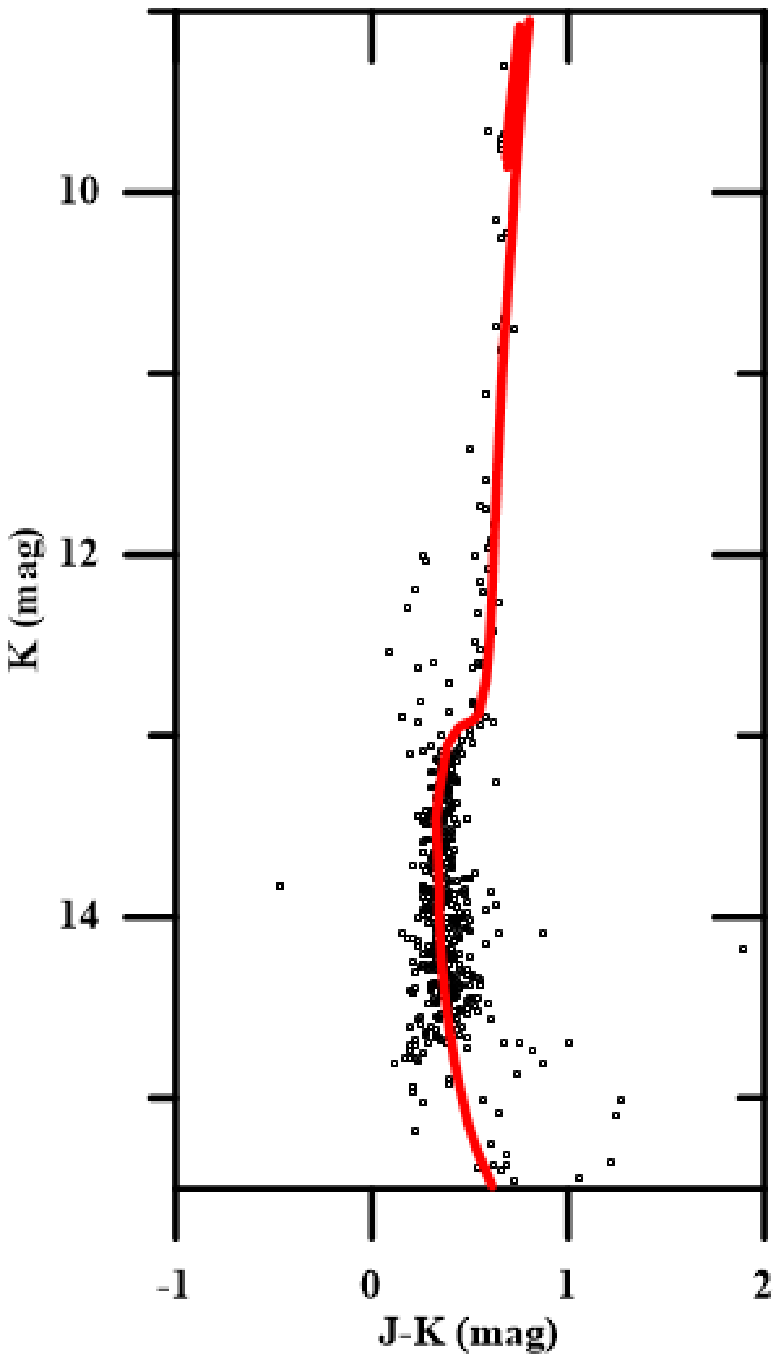}
\caption{Theoretical Padova isochrones with Z=0.019 and log(age) = 9.85$\pm$0.05 fitted over
($J,J-H$) and ($K,J-K$) isochrones for NGC 188.}
\label{iso}
\end{figure}

Many photometric parameters including reddening and distance modulus
can be determined by fitting theoretical isochrones to the observed CMDs.
Bonatto et al. (2004) fitted Padova isochrones to CMDs in ($J,J-H$) and ($K,J-K$)
bands. We used the equations given by Carpenter (2001) to convert
$K_s$ magnitudes to $K$ magnitudes. The fitting of isochrones with the observed CMDs are 
shown in Fig.~\ref{iso}. The visual best fit solar metallicity ($Z$=0.019) isochrones 
provide the cluster age as log(age) = 9.85. 

Reddening of the cluster has been determined using Schlegel et al. (1998) 
and Schlafly et al. (2011). We have the coefficient ratios $A_J/A_V$ = 0.276 
and $A_H/A_V$ = 0.176 , which are derived using absorption ratios by 
Schlegel et al. (1998), while the ratio
$A_{K_s}/A_V$ = 0.118 was derived by Dutra et al. (2002). Here we have the following
results for the color excess of 2MASS photometric system by Fiorucci $\&$ Munari
(2003): $E_{J-H}/E_{B-V}$ = 0.309$\pm$ 0.130, $E_{J-K}/E_{B-V}$ = 0.485$\pm$ 0.150;
where $R_V = A_V/E_{B-V}$ = 3.1. We used these formulae for the cluster under study to 
correct CMDs from the effect of reddening, i.e. $A_{K}/E_{B-V}$ = 0.365 and
$A_J/E_{B-V}$ = 0.879. Conclusively, the distance of the cluster
from the Sun can be calculated. We estimated the distance as 
1721$\pm$41 pc, which we will use to study the kinematics of the cluster in Sect.~\ref{kin}.
This distance can also be used to determine the cluster's distance 
to the Galactic center ($R_{gc}$, the projected 
distance to the Galactic plane ($X_{\odot}, Y_{\odot}$)
and the distance from the Galactic plane $Z_{\odot}$ (Tadross 2012).

\subsection{Luminosity and mass function}
\begin{figure}[]
\centering
\vspace{-0.7cm}
\includegraphics[width=8cm,height=8cm]{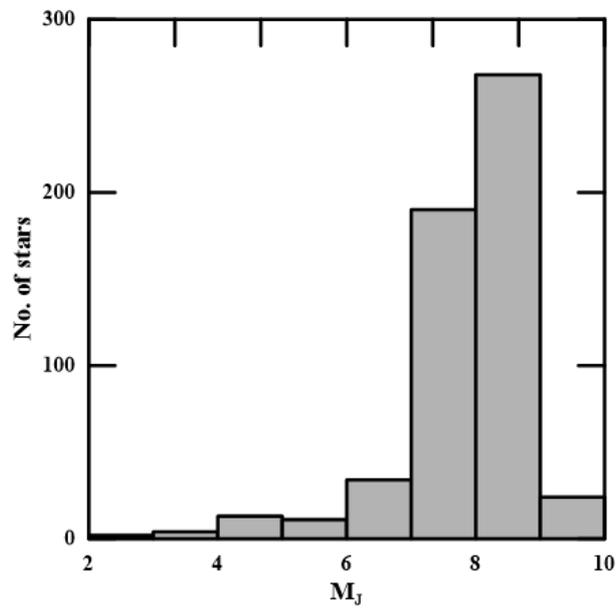}
\caption{The apparent LF of NGC 188.}
\label{lf}
\end{figure}

One of the main attributes of studying open clusters is to study mass function (MF)
and check how it varies with different star forming conditions and stellar evolution.
Mass function is derived using luminosity function (LF), which is relative number of
stars in certain interval bins of absolute magnitudes. Mass function is related 
to the LF by a relation called mass-luminosity relation (MLR). 
LF for NGC188 is plotted in Fig.~\ref{lf}.
Scalo (1986) defined the initial mass function as an empirical relation 
that describes the mass distribution (i.e. histogram of stellar masses) 
of a population of stars in terms of their theoretical initial mass 
(the mass they were formed with) which is mathematically defined as:
\begin{center}
$\frac{dN}{dM} \propto M^{-{\alpha}} $
\end{center}
where $dN/dM$ represents the number of stars in the mass interval ($M$, $M+DM$)
and $\alpha$ is a dimensionless exponent.

Salpeter (1955) established the IMF for massive stars ($>1 M_{\odot}$) as $\alpha$=2.35.
The steep slope of IMF indicates that during the star formation in the cluster, the number of 
low-mass stars is greater than high-mass stars. MLR of NGC 188 could be constructed
using the adopted isochrones (Bonatto et al. 2004). The relation is a polynomial
function of second order, i.e.:
\begin{center}
$\frac{M}{M_\odot} = 1.617385326-0.2204258896 M_K + 0.004586772674 {M^2_ K} $
\end{center}

\begin{figure}[]
\centering
\vspace{-0.7cm}
\includegraphics[width=8cm,height=8cm] {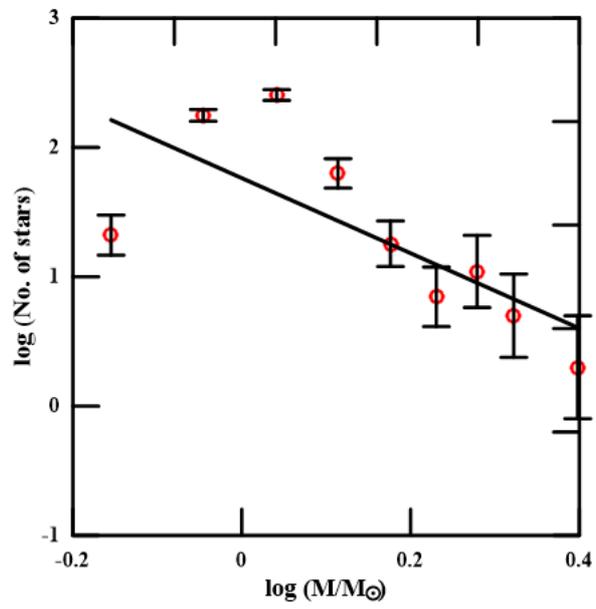}
\caption{Mass function of open cluster NGC 188. The slope of MF is calculated as -2.90$\pm$0.21.}
\label{mf}
\end{figure}

The MF is shown in Fig.~\ref{mf} and the MF slope for NGC 188
is determined as 2.90$\pm$0.21, which is in good agreement with the Salpeter value. 

\subsection{Dynamical relaxation time}

The relaxation time $T_{relax}$ is the time-scale on which the cluster will lose
all traces of its initial condition (Yadav et al. 2013).
Due to internal dynamics of the cluster,
the contraction and destruction forces make the cluster
approach a Maxwellian equilibrium. Through this time, low mass stars may
evaporate with large random velocities trying to occupy a larger volume than the high
mass stars do (Mathieu and Latham 1986). Mathematically, relaxation time has the
following form (Spitzer and Hart 1971):
\begin{center}
\begin{displaymath}
\hspace{2.0cm}T_{relax} = \frac {8.9 \times 10^{5} N^{1/2} R_{h}^{3/2}}{ <m>^{1/2}log(0.4N)}
\end{displaymath}
\end{center}
where  $R$$_{h}$ is the radius containing half of the cluster mass, 
$N$ is the number of cluster members and
$<m>$ is the average mass of the cluster stars.
Using the above formula, we can calculate the dynamical relaxation time
for NGC 188. Also, the dynamical evolution parameter $\tau$ can be calculated for the
cluster using the formula:
\begin{center}
$\tau = \frac{age}{T_{relax}}$.
\end{center}
If $\tau >>$ 1, then the cluster may be called dynamically relaxed and vice versa.

Jeffries et al. (2001) gives equation of tidal radius $r_t(pc)$, which depend on the 
total mass $M_c$ of the cluster, i.e.
\begin{center}
$r_t = 1.46 \sqrt[3]{M_c}$. 
\end{center}
Also, Peterson $\&$ King (1975), define the cluster 
concentration parameter as:
\begin{center}
$C = log(\frac{r_{lim}}{r_{core}})$.
\end{center}

Table 1 presents the results of our work devoted to the photometric analysis
of NGC 188 and its comparison with other published works. 

\section{Kinematical analysis}
\label{kin}
\subsection{The vertex ($A_0,D_0$) of a moving cluster}
 
\begin{figure}[]
\centering
\vspace{-0.7cm}
\includegraphics[width=13cm,height=13cm] {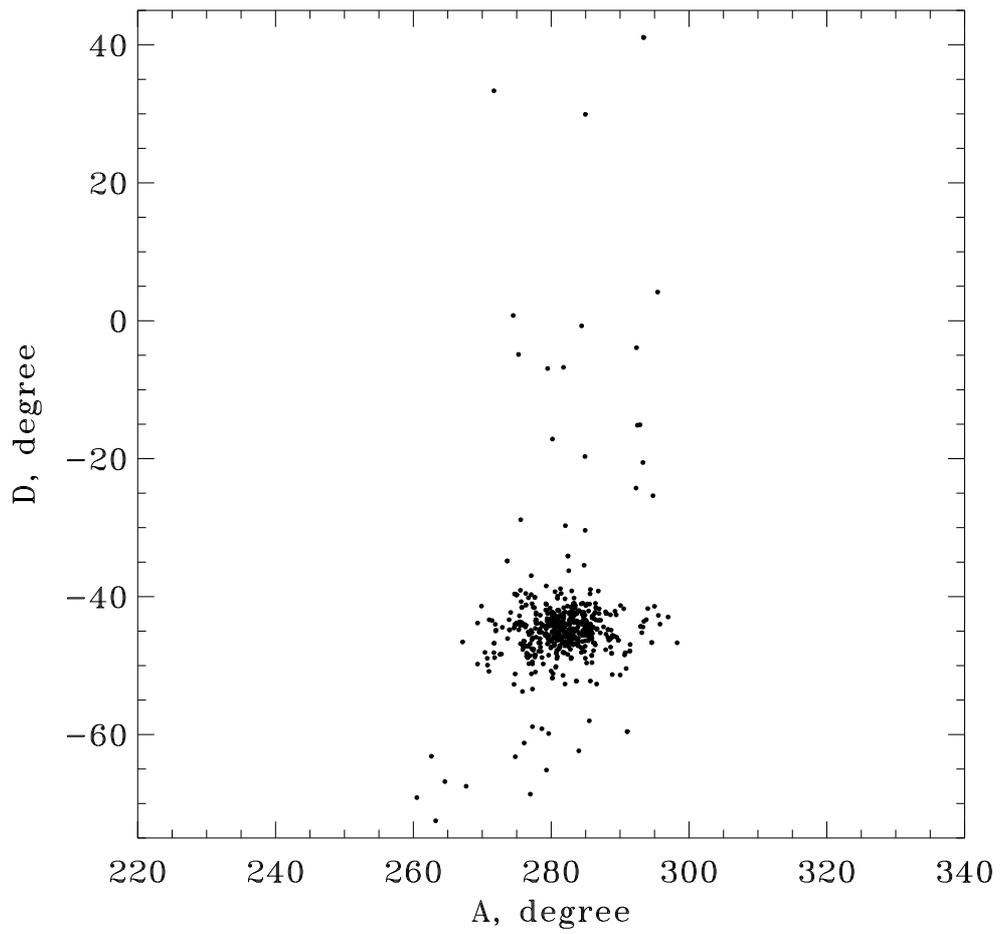}
\caption{The AD-diagram for 562 stars with known radial velocities and having membership probability $>$50\%}
\label{ad}
\end{figure}

In order to compute the vetrtex using the most probable cluster 
members selected from Platais et al. (2003), we used apex diagram method
(AD-method). This method has been used and explained in detail by
Chupina et al. (2001, 2006) and Vereshchagin et al (2014).
AD-diagram, plotted using individual apexes of stars represents
the distribution of stars in the equatorial coordinate system.
The coordinates are obtained from the solution of a geometrical problem
in which the coordinates of the points of intersection of the vectors of
spatial velocities of stars are displaced observation points with the
celestial sphere. Figure~\ref{ad} shows the AD-diagram of individual apexes
of stars in the cluster region.

We followed the computation algorithm explained in Elsanhoury (2015).
We present a brief preview of the process here. 
Let us consider a group of $N_i$ cluster stars at a distance $r_i$ (pc) 
with coordinates ($\alpha,\delta$), which are moving with proper motion $\mu$ (mas/yr)
and radial velocity $V_r$ (km/sec).
We used the distance value of the stars in the stellar group (cluster) under consideration
as $r_i=1721\pm41 pc$, as calculated in Sect.~\ref{phot}. 

The velocity components $(V_x, V_y, V_z)$ along $x,y$ and $z$ axes
of a coordinate system centered at the Sun can be given
by the formulae given by Smart (1958):
\begin{center}
%\begin{displaymath}
$V_x = -4.74r_i\mu_\alpha\cos\delta\sin\alpha-4.74r_i\mu_\delta\sin\delta\cos\alpha+V_r\cos\delta\cos\alpha$ \\
$V_y = +4.74r_i\mu_\alpha\cos\delta\cos\alpha-4.74r_i\mu_\delta\sin\delta\sin\alpha+V_r\cos\delta\sin\alpha$ \\
$V_z = +4.74r_i\mu_\delta\cos\delta+V_r\sin\delta $
%\end{displaymath}
\end{center}

After averaging the values of $(V_x, V_y, V_z)$ of different stars, we calculate the 
equatorial coordinates of the convergent point, as discussed in 
Chupina et al. (2001), i.e.:
\begin{center}
$A_0 = \tan^{- 1}(\frac{\overline{V_y}}{\overline{V_x}}) $ \\
$D_0 = \tan^{- 1}(\frac{\overline{V_z}}{\sqrt{\overline{V^2_x}+\overline{V^2_y}}}) $
\end{center}

\subsection{The cluster velocity}

The velocity of the cluster can be calculated by the following formula:
\begin{center}
$V = \frac{\sum_{i=1}^{N} V_r^{(i)}\cos\lambda_i}{\sum_{i=1}^{N}\cos^2\lambda_i}, $
\end{center}
where $\lambda$ is the angular distance from the star to apex defined as:
\begin{center}
$\lambda_i = \cos^{- 1}[\sin\delta_i\sin D_0+\cos\delta_i\cos D_0\cos(A_0-\alpha_i)].   $
\end{center}

\subsection{The center of the cluster}

The center of the cluster can be derived by the simple method of finding the 
equatorial coordinates of the center of mass for $N_i$ number of discrete objects, as following:
\begin{center}
$x_c = \frac {[\sum_{i=1}^{N} r_i\cos\alpha_i\cos\delta_i]}{N}, $\\
$y_c = \frac {[\sum_{i=1}^{N} r_i\sin\alpha_i\cos\delta_i]}{N}, $\\
$z_c = \frac {[\sum_{i=1}^{N} r_i\sin\delta_i]}{N}.$
\end{center}

\subsection{The components of space velocity}

In order to compute components of space velocity in the Galactic
space coordinates $(U,V,W)$, we use the tranformations given by Murray (1989).
The formulae are given below:
\begin{center}
$U = -0.054875539V_x-0.873437105V_y-0.483834992V_z , $ \\
$V =  0.494109454V_x-0.444829594V_y+0.746982249V_z , $ \\
$W = -0.867666136V_x-0.198076390V_y+0.455983795V_z. $ 
\end{center}
While the mean velocities are calculates as:
\begin{center}
$\overline{U} = \frac{\sum_{i=1}^{N} U_i}{N} , $\\
$\overline{V} = \frac{\sum_{i=1}^{N} V_i}{N} , $\\
$\overline{W} = \frac{\sum_{i=1}^{N} W_i}{N}.  $\\
\end{center}

\subsection{Elements of Solar motion}

The Solar motion can be defined as the absolute value of the 
velocity of the Sun relative to the group of stars under consideration, i.e.:
\begin{center}
$S_{\odot} = \sqrt{\overline{U}^2+\overline{V}^2+\overline{W}^2}   (km/sec). $
\end{center}

The Galactic longitude ($l_A$) and Galactic latitude ($b_A$)
of the Solar apex are determined as:

\begin{center}
$l_A = \tan^{-1}(\frac{-\overline{V}}{\overline{U}}) , $\\
$b_A = \sin^{-1}(\frac{-\overline{W}}{S_{\odot}}). $\\
\end{center}

The above three parameters $S_{\odot}$, $l_A$ and $b_A$
are called the elements of Solar motion with respect to the
the group of stars under consideration.

Table 2 presents the results of the kinematical
studies conducted in this article.
%%%%%%%%%%%%%%%%%%%%%%%%%%%%%%%%%%%%%%%%%%%%%%%%%%%%%%%%%%%%%%%%%%%%%%%%%%%%%%%%
\section{Conclusions}
\label{con}

In the present study, we extracted photometric ($J,H,K$) data 
with proper motions and radial velocities from various data sources
including 2MASS catalogue. The main conclusions of the present photometric 
and kinematic study of open cluster NGC 188 can be given as follows:

\begin{enumerate}

\item We have drawn the radial density profile of
the cluster and determined the limiting radius and core radius according to
King (1966) model. We present a CMD using most probable cluster members and 
determined various fundamental parameters like
distance, age and reddening after correcting for interstellar reddening. 
The cluster's age was derived as 7.08$\pm$0.04 Gyr using near-IR magnitude data. 

\item We plotted the luminosity function and mass function diagrams for NGC 188.
LF show a gradual increase towards low luminosity stars from high luminosity ones. 
The mass function slope, a dimensionless exponent was determined as $\alpha =$ 2.90$\pm$0.21,
which is in good agreement with Salpeter's value. The dynamical evolution parameter
was found to be $\tau >> 1$, which indicates that the cluster is dynamically relaxed.

\item In the kinematical studies of the cluster, we computed the apex of the cluster by
AD-diagram method as $A_0= 281^\circ$.88, $D_0=-44^\circ$.76. 
Then, we followed an algorithm to compute velocity,
cluster center, space velocity and elements of solar motion for the cluster.
\end{enumerate}

\section{Acknowledgments}
We are thankful to an anonymous referee for careful reading of the paper
and constructive comments. 
N. V. Chupina and S. V. Vereshchagin are partly supported by
the Russian Foundation for Basic Research (RFBR, grant number is 16-52-12027).
Devesh P. Sariya and Ing-Guey Jiang acknowledge the grant from 
Ministry of Science and Technology
(MOST), Taiwan. The grant numbers are MOST 103-2112-M-007-020-MY3 and
NSC 100-2112-M-007-003-MY3.
This publication makes use of data products from the Two Micron All Sky Survey, 
which is a joint project of the University of Massachusetts and 
the Infrared Processing and Analysis Center/California Institute of Technology, 
funded by the National Aeronautics and Space Administration and the National Science Foundation.
%\bliographystyle{model2-names}

%%%%%%%%%%%%%%%%%%%%%%%%%%%%%%%%%%%%%%%%%%%%%%%%%%%%%%%%%%%%%%%%%%%%%%%%%%%%%%%%%%%%%%%%%%%%%%%%%%%%%%%%%%
\clearpage
%%%%%%%%%%%%%%%%%%%%%%%%%%%%%%%%%%%%%%%%%%%%%%%%%%%%%%%%%%%%%%%%%%%%%%%%%%%%%%%%%%%%%%%%%%%%%%%%%%%%%%%%%%%
%%%%%%%%%%%%%%%%%%%%%%%%%%%%%%%%%%%%%%%%%%%%%%%%%%%%%%%%%%%%%%%%%%%%%%%%%%%%%%%%%%%%%%%%%%%%%%%%%%%%%%%%%%
\clearpage
%%%%%%%%%%%%%%%%%%%%%%%%%%%%%%%%%%%%%%%%%%%%%%%%%%%%%%%%%%%%%%%%%%%%%%%%%%%%%%%%%%%%%%%%%%%%%%%%%%%%%%%%%%
%%%%%%%%%%%%%%%%%%%%%%%%%%%%%%%%%%%%%%%%%%%%%%%%%%%%%%%%%%%%%%%%%%%%%%%%%%%%%%%%%%%%%%%%%%%%%
\clearpage
%%%%%%%%%%%%%%%%%%%%%%%%%%%%%%%%%%%%%%%%%%%%%%%%%%%%%%%%%%%%%%%%%%%%%%%%%%%%%%%%%%%%%%%%%%%%
%%%%%%%%%%%%%%%%%%%%%%%%%%%%%%%%%%%%%%%%%%%%%%%%%%%%%%%%%%%%%%%%%%%%%%%%%%%%%%%%%%%%%%%%%%%%
\clearpage
%%%%%%%%%%%%%%%%%%%%%%%%%%%%%%%%%%%%%%%%%%%%%%%%%%%%%%%%%%%%%%%%%%%%%%%%%%%%%%%%%%%%%%%%%%%%
%%%%%%%%%%%%%%%%%%%%%%%%%%%%%%%%%%%%%%%%%%%%%%%%%%%%%%%%%%%%%%%%%%%%%%%%%%%%%%%%%%%%%%%%%%%%
\clearpage
%%%%%%%%%%%%%%%%%%%%%%%%%%%%%%%%%%%%%%%%%%%%%%%%%%%%%%%%%%%%%%%%%%%%%%%%%%%%%%%%%%%%%%%%%%%%
%%%%%%%%%%%%%%%%%%%%%%%%%%%%%%%%%%%%%%%%%%%%%%%%%%%%%%%%%%%%%%%%%%%%%%%%%%%%%%%%%%%%%%%%%%%%%%%
\clearpage
\begin{table*}
\small
\caption{The results of photometric analysis for NGC 188.}
\vspace{0.5cm}
\centering
\begin{tabular}{lll|lll}
\hline\hline
Parameters&Value &Reference&Parameters&Value &Reference\\
\hline
Age$(Gyr)$        &7.08$\pm$0.04   &Present work                     &$E(J-K)$          &0.021            &Present work\\
                  &7.0$\pm$0.5     &Jiaxin et al. (2015)             &$E(J-H)$          &0.007            &Present work\\
                  &6               &Chumak et al. (2010)             &                  &0.0              &Bonatto et al. (2005)\\
                  &6.2$\pm$0.2     &Meibom et al. (2009)             &Distance$(pc)$    &1721$\pm$41      &Present work\\
                  &7.0$\pm$0.5     &Krusberg \& Chaboyer (2006)      &                  &1714             &Jiaxin et al. (2015)\\
                  &7.0$\pm$0.7     &Fornal et al. (2007)             &                  &1650$\pm$50      &Chumak et al. (2010)\\
                  &7.0$\pm$1.0     &Bonatto et al. (2005)            &                  &1770$\pm$75      &Meibom et al. (2009)\\
                  &6.8$\pm$0.7     &VandenBerg \& Stetson (2004)     &                  &1700$\pm$100     &Fornal et al. (2007)\\
                  &6.4             &Michaud et al. (2004)            &                  &1660$\pm$80      &Bonatto et al. (2005)\\
                  &7.0             &Sarajedini (1999)                &                  &1900             &Sarajedini (1999)\\
                  &7.0             &von Hippel \& Sarajedini (1998)  &$f_0 (pc)$        &4.38$\pm$0.19    &Present work\\
No. of members    &562             &Present work                     &$r_{lim}(pc)$     &6.23$\pm$0.65    &Present work\\
Metallicity $(Z)$ &0.019           &Present work                     &$R_h(pc)$         &3.12$\pm$0.33    &Present work\\
                  &0.020           &Jiaxin et al. (2015)             &                  &4.0              &Chumak et al. (2010)\\
                  &0.015           &VandenBerg \& Stetson (2004)     &$R_{core}(pc)$    &1.91             &Present work\\
$(m-M)_0$         &11.188$\pm$0.06 &Present work                     &                  &1.89             &Jiaxin et al. (2015)\\
                  &11.17$\pm$0.08  &Jiaxin et al. (2015)             &                  &1.3$\pm$0.1      &Bonatto et al. (2005)\\
                  &11.35           &Chumak et al. (2010)             &                  &2.1              &Chumak et al. (2010)\\
                  &11.24$\pm$0.09  &Meibom et al. (2009)             &$C$               &0.55             &Present work\\
                  &11.23$\pm$0.14  &Fornal et al. (2007)             &                  &1.07             &Jiaxin et al. (2015)\\
                  &11.1$\pm$0.1    &Bonatto et al. (2005)            &                  &1.21             &Chumak et al. (2010)\\
                  &11.44           &von Hippel \& Sarajedini (1998)  &                  &1.2$\pm$0.1      &Bonatto et al. (2005)\\
                  &11.44           &Sarajedini (1999)                &$M_C (M_{\odot})$ &616.5            &Present work\\
$E(B-V)$          &0.033$\pm$0.03  &Present work                     &$<m>(M_{\odot})$  &1.097            &Present work\\
                  &0.036$\pm$0.01  &Jiaxin et al. (2015)             &$r_t(pc)$         &12.43            &Present work\\
                  &0.083           &Chumak et al. (2010)             &                  &22.3             &Jiaxin et al. (2015)\\
                  &0.087           &Meibom et al. (2009)             &                  &21$\pm$4         &Bonatto et al. (2005)\\
                  &0.025$\pm$0.005 &Fornal et al. (2007)             &                  &34               &Chumak et al. (2010)\\
                  &0.09            &von Hippel$\&$ Sarajedini (1998) &$X_{\odot} (pc)$  &$-$863.041       &Present work\\
                  &0.09            &Sarajedini (1999)                &$Y_{\odot} (pc)$  &1336.97          &Present work\\
                  &0.0             &Bonatto et al. (2005)            &$Z_{\odot} (pc)$  &655.378          &Present work\\
                  &0.087           &VandenBerg \& Stetson (2004)     &$\alpha$          &2.90$\pm$0.21    &Present work\\
$T_{relax} (Myr)$ &47.09           &Present work                     &                  &2.15             &Jiaxin et al. (2015)\\
                  &64              &van den Bergh \& Sher (1960)     &  $R_{gc} (pc)$   &8672.48$\pm$74.3 &Present work\\
$\tau$            &150.35          &Present work                     &                  &8900$\pm$100     &Bonatto et al. (2005)\\
\hline
\label{tab1}
\end{tabular}
\end{table*}
%
%%%%%%%%%%%%%%%%%%%%%%%%%%%%%%%%%%%%%%%%%%%%%%%%%%%%%%%%%%%%%%%%%%%%%%%%%%%%%%%%%%%%%%%%%%%%%%%
%%%%%%%%%%%%%%%%%%%%%%%%%%%%%%%%%%%%%%%%%%%%%%%%%%%%%%%%%%%%%%%%%%%%%%%%%%%%%%%%%%%%%%%%%%%%%%%
\clearpage
\begin{table*}
\caption{The results of kinematical analysis of NGC 188.}
\vspace{0.5cm}
\centering
\begin{tabular}{lll}
\hline\hline
Parameters&Value &Reference\\
\hline
\noalign{\smallskip}
$(\overline{V_x},\overline{V_y}, \overline{V_z}) (km/sec)$  &$(8.93,-42.46,-43.02)$               &Present work          \\
$(A_0,D_0)$                                                 &($281^\circ$.88, $-44^\circ$.76)     &Present work          \\
$V (km/sec)$                                                &$-42.87\pm0.3$                       &Present work          \\
                                                            &$-42.35\pm0.05$                      &Xin-Hua Gao (2014)    \\
                                                            &$-42.36\pm0.04$                      &Geller et al. (2008)  \\
                                                            &$-42.4\pm0.1  $                      &Chumak et al. (2010)\\
$(x_c, y_c, z_c) (pc)$                                      &$(139.46,29.3,1715.08)$              &Present work          \\
$(\overline{U},\overline{V}, \overline{W}) (km/sec)$        &$(57.41,-8.83,-18.96)$               &Present work          \\
$S_{\odot} (km/sec)$                                        &61.1                                 &Present work          \\
$l_A$                                                       &$8^\circ$.745                        &Present work          \\
$b_A$                                                       &$18^\circ$.08                        &Present work          \\
\hline
\label{tab2}
\end{tabular}
\end{table*}
%%%%%%%%%%%%%%%%%%%%%%%%%%%%%%%%%%%%%%%%%%%%%%%%%%%%%%%%%%%%%%%%%%%%%%%%%%%%%%%%%%%%%%%%%%%%%%%
\end{document}